\documentclass[final]{aipproc}

\layoutstyle{8x11double}
\usepackage[colorlinks=true,citecolor=blue]{hyperref}
\usepackage{amsmath,amssymb,psfrag}
\usepackage{multirow}
\usepackage{dcolumn}
\usepackage{xspace}

\def\WMAP{{\small WMAP}\xspace}
\def\EWSB{{\small EWSB}\xspace}
\def\MPlanck{\ensuremath{M_{\rm P}}\xspace}
\def\M2{\ensuremath{M_{2}}\xspace}
\def\m0{\ensuremath{m_{0}}\xspace}
\def\SMgr{SU(3)$\times$SU(2)$\times$U(1)\xspace}
\def\SU51{SU(5)$\times$U(1)\xspace}
\def\SUPS{SU(4)$\times$SU(2)$\times$SU(2)\xspace}
\def\SUSO7{SU(2)$\times$SO(7)\xspace}
\def\hc{\textrm{\small H.C.}}

\def\vev{\emph{vev}}
\def\lsim{\mathrel{\rlap{\raise 2.5pt \hbox{$<$}}\lower 2.5pt\hbox{$\sim$}}}
\def\gsim{\mathrel{\rlap{\raise 2.5pt \hbox{$>$}}\lower 2.5pt\hbox{$\sim$}}}

\def\GUT{\ensuremath{\text{\sc gut}}}
\def\EW{\ensuremath{\text{\sc ew}}}

\newcommand{\refeq}[1]{Eq.~\eqref{#1}}
\newcommand{\Neu}[1]{\widetilde \chi_{#1}^0}
\newcommand{\Cha}[1]{\widetilde \chi_{#1}^\pm}
\newcommand{\wbar}[1]{\mkern1mu\overline{\mkern-3mu#1\mkern-1mu}\mkern2mu}
\RequirePackage{fancyhdr} 
\newcommand{\preprintJL}[1]{\thispagestyle{fancy}
  \renewcommand{\headrulewidth}{0pt} \fancyhead[R]{#1}
}
\newcommand{\forceemail}[1]{
  \let\thefootnote\relax\footnotetext{\emph{e-mail}: #1}
}

\begin{document}

\title{Relic density in SO(10) SUSY GUT models}


\author{Katri Huitu}{
  address={Department of Physics, and 
    Helsinki Institute of Physics, \\P.O.~Box 64,
    {FIN-00014} University of Helsinki, Finland}
}

\author{Jari Laamanen}{
  address={Institut f\"ur Physik, 
    Technische Universit\"at Dortmund, {D-44221} Dortmund, Germany}
}

\begin{abstract}
  Non-universal boundary conditions in grand unified theories can lead
  to non-universal gaugino masses at the unification scale.  In
  $R$-parity preserving theories the lightest supersymmetric particle
  is a natural candidate for the dark matter. In this talk the
  composition of the lightest neutralino is studied, when nonuniversal 
  gaugino masses come from representations of SO(10).  In these cases,
  the thermal relic density compatible with \WMAP\ observations is found.
\end{abstract}

\maketitle

\preprintJL{DO-TH-08/06 \\ HIP-2008-30/TH}
\forceemail{jari.laamanen@uni-dortmund.de}


\section{Introduction}
\label{intro}

The phenomenology of supersymmetric models depends crucially on the
compositions of neutralinos and charginos.  In addition to the
laboratory studies, relevant input is obtained from the dark matter
searches, where the \WMAP\ satellite has put precise limits on the
relic density. Supersymmetric theories which preserve $R$-parity
contain a natural candidate for the cold dark matter particle.  If the
lightest neutralino is the lightest supersymmetric particle (LSP), it
can provide the appropriate relic density.

In many supergravity type models the lightest neutralino is bino-like,
which often leads to too high thermal relic density, as compared to
the limits provided by the \WMAP\ experiment. When the gaugino masses
are not universal at the grand unification scale, the resulting
neutralino composition changes from the usual universal gaugino mass
case \cite{Huitu:2005wh}.

In this talk, the thermal relic density of the neutralino LSP is
studied, when gaugino masses are due to nonuniversal representations
of SO(10) grand unified theory (GUT)
\cite{Fritzsch:1974nn,Zhao:1981me}.  SO(10) has many attractive
features when compared to SU(5), in which the relic density has been
studied e.g. in \cite{Huitu:2008sa}: the one family matter fermions
can be put to one spinor representation of SO(10), including the right
handed neutrino \cite{Mohapatra:1979ia}, representations are anomaly
free, and $R$-parity may result from some gauge symmetry breaking, to
mention some.


Gaugino masses originate from the non-renormalizable terms in the
$N=1$ supergravity Lagrangian involving the gauge kinetic function
$f_{ab}(\Phi)$ \cite{Cremmer:1982wb}. The gauge part of the Lagrangian
contains the gauge kinetic function coupling with two field strength
superfields $W^a$.
The function $f_{ab}(\Phi)$ is an analytic function of the chiral
superfields $\Phi$ in the theory. It transforms as a symmetric product
of two adjoint representations under the GUT gauge group, since the
field strength superfield transforms in the adjoint representation.
In order to generate a mass term for the gauginos, the gauge kinetic
function must be non-minimal, i.e. it must not be a constant
\cite{Ferrara:1982qs}.
The chiral superfields $\Phi$ consist of a set of gauge singlet
superfields $\Phi^s$ and gauge nonsinglet superfields $\Phi^n$,
respectively, under the grand unified group. If the auxiliary part
$F_\Phi$ of a chiral superfield $\Phi$ in the $f_{ab}(\Phi)$ gets a
\vev, then gaugino masses arise from the coupling of $f_{ab}(\Phi)$
with the field strength superfield $W^a$. The Lagrangian for the
coupling of gauge kinetic function with the gauge field strength is
written as
\begin{eqnarray}
  {\cal L}_{gk} = \int d^2 \theta f_{ab}(\Phi) W^a W^b + \hc,
  \label{gauge-fs-kinetic}
\end{eqnarray} 
where $a$ and $b$ are gauge group indices (for example, $a,b =
1,2,...,45$ for ${\rm SO(10)}$), and repeated indices are summed
over. The scalar fields in $f_{ab}(\Phi)$ are suppressed by the
inverse powers of \MPlanck, so the gauge kinetic
function $f_{ab}(\Phi)$ can be expanded,
\begin{eqnarray}
  f_{ab}(\Phi) = f_0(\Phi^s)\delta_{ab} + \sum_n f_n(\Phi^s)
  \frac{\Phi^n_{ab}}{M_P} + \cdots,
  \label{gauge-kinetic}
\end{eqnarray}
where $\Phi^s$ and $\Phi^n$ are the singlet and nonsinglet chiral
superfields, respectively. Here $f_0(\Phi^s)$ and $f_n(\Phi^s)$ are
functions of gauge singlet superfields $\Phi^s$, and $M_P$ is some
large scale. When $F_\Phi$ gets a \vev\ $\langle F_\Phi \rangle$, the
interaction (\ref{gauge-fs-kinetic}) gives rise to gaugino masses:
\begin{eqnarray}
  {\cal L}_{gk} \supset \frac{{\langle F_\Phi \rangle}_{ab}}
  {M_P}\lambda^a \lambda^b + \hc,
  \label{gmassterm}
\end{eqnarray}
where $\lambda^{a,b}$ are gaugino fields. The non-universal gaugino
masses are generated by the nonsinglet chiral superfield $\Phi^n$ that
appears linearly in the gauge kinetic function $f_{ab}(\Phi)$
\eqref{gauge-kinetic}.


Gauginos belong to the adjoint representation of the gauge group,
which in the case of ${\rm SO(10)}$ is the {\bf 45} dimensional
representation.  Because \refeq{gmassterm} must be gauge invariant,
$\Phi$ and $F_\Phi$ must belong to some of the following
representations appearing in the symmetric product of the two {\bf 45}
dimensional representations of ${\rm SO(10)}$ \cite{Chamoun:2001in}:
\begin{eqnarray}
  ({\bf 45 \otimes 45})_{Symm} = 
  {\bf 1 \oplus 54 \oplus 210 \oplus 770}.
   \label{symmprod}
\end{eqnarray}
The representations {\bf 54}, {\bf 210} and {\bf 770} may lead to
non-universal gaugino masses, while the {\bf 1} dimensional
representation gives manifestly the universal gaugino masses.  The
relations between the gaugino masses are determined by the
representation invariants, and are specific for each of the
representations. Because the gauge kinetic function
(\ref{gauge-kinetic}) can get contributions from several different
$\Phi$'s, a linear combination of any of the representations is also
possible. In that case the gaugino masses are not anymore uniquely
determined, in contrast to the contribution purely from one
representation. Because of this it is more instructional to study the
representations separately. In other words, we assume that the
dominant component of the gaugino masses comes only from one
representation, and more specifically, since we want to study
non-universal gaugino masses, from one of the non-singlet
representation.

\thispagestyle{plain} 

\section{Dark matter in SO(10) representations}
\label{sec:rd}

\subsection{Breaking Chains: SO(10) \texorpdfstring{$\to
    \boldsymbol{H}\to$}{-- H --} SM}

The GUT group SO(10) breaks down to the Standard Model (SM) gauge
group \SMgr via some intermediate gauge group $H$. Therefore the
gaugino mass relations depend also on the gauge group breaking chain,
in addition to the representation invariants coming from the gauge
kinetic function. Moreover, the intermediate breaking scale affects
also the generated gaugino masses. However, if the gauge breaking
from the GUT scale to the SM scale takes place all at the GUT scale,
these loop-induced messenger contributions \cite{Giudice:1997ni} can
be neglected in comparison to the tree-level contributions.
Table \ref{tab:chains} shows various possible SO(10) breaking chains
\cite{Aulakh:1982sw,Chamoun:2001in} for the two chosen
representations.
Some of the subgroups lead to universal gaugino masses, or to massless
gauginos.
In this talk we suppose that the gauge breaking from SO(10) to \SMgr
happens at the GUT scale, and that the GUT breaking does not affect
the gauge coupling unification. We also limit our study to the
representations {\bf 54} and {\bf 210} and to three specific
intermediate gauge groups, \SUPS, \SUSO7 and \SU51.
\begin{table}[htb]
\setlength{\tabcolsep}{3pt}
  \caption{\label{tab:chains} Various breaking chains of SO(10)}
  \begin{tabular}{ccc}
    \hline 
    $F_\Phi$ & $H$ & Subgroup description  \\
    \hline
    \multirow{3}{*}{\bf 54} &
    $\scriptstyle{\text{SU}(4)\times \text{SU}(2)
      \times \text{SU}(2)}$ & {\small Pati--Salam}\\
    & $\scriptstyle{ \text{SU}(2)\times \text{SO}(7)}$ & \\
    & $\scriptstyle{\text{SO(9)}}$ & {\small Universal gauginos}\\
    \hline
    \multirow{4}{*}{\bf 210} &
    $\scriptstyle{\text{SU}(4)\times \text{SU}(2)\times \text{SU}(2)}$ & 
    {\small Massless gluino}\\
    & $\scriptstyle{\text{SU}(3)\times \text{SU}(2)
      \times \text{SU}(2) \times \text{U}(1)}$ & 
    {\small Massless }$\scriptstyle{\text{SU}(2)_\text{L}}${ \small gauginos}\\
    & $\scriptstyle{\text{SU}(3)\times \text{SU}(2)
      \times \text{U}(1) \times \text{U}(1)}$ & \\
    & $\scriptstyle{ \text{SU}(5)\times \text{U}(1)}$ & {\small
      'Flipped' }$\scriptstyle{\text{SU}(5)}$\\
    \hline
  \end{tabular}
\end{table}
Table \ref{tab:gaug} displays the ratios of resulting gaugino masses
at tree level as they arise when $F_\Phi$ belongs to various
representations of ${\rm SO(10)}$. The relations at the electroweak
scale resulting from 1-loop renormalization group running are also
displayed.
\begin{table}[htb]
\setlength{\tabcolsep}{1pt}
  \caption{\label{tab:gaug} Ratios of the gaugino masses at the GUT
    scale in the normalization $M_3^\GUT \equiv  {M_3}(\GUT)$ = 1, and at the electroweak
    scale in the normalization $M_3^\EW \equiv {M_3}(\EW) = 1$} 
  \centering
  \begin{tabular}{cccccr@{\hspace{5pt}}cc}
    \hline
    $F_\Phi$ & $H$ & $M_1^\GUT$ & $M_2^\GUT$ & $M_3^\GUT$ & 
    $M_1^{\EW}$ & $M_2^{\EW}$ & $M_3^{\EW}$
    \\ \hline 
    {\bf 1} & & 1 & 1 & 1 & { 0.14} & 0.29 & 1 \\
    {\bf 54} &$\scriptstyle{\text{SU}(4)\times \text{SU}(2)\times \text{SU}(2)}$ 
    & -1\hspace{3pt} & -1.5 & 1 & { -0.15} & -0.44 & 1 \\
    {\bf 54} &$\scriptstyle{\text{SU}(2)\times \text{SO}(7)}$ 
    & 1 & -7/3 & 1 & { 0.15} & {-0.68} & 1 \\
    {\bf 210} &$\scriptstyle{\text{SU}(5)\times \text{U}(1)}$ 
    & -96/25 & 1 & 1 &-0.56 & { 0.29} & 1 \\
    \hline
  \end{tabular}
\end{table}

\subsection{Representation 210}
\label{subsec:rep210}
\begin{figure} 
  \centering
  \psfrag{m2}{$M_{2}$}
  \psfrag{m0}{$m_{0}$}
  \psfrag{lsp}{\hspace{-6.6mm} \footnotesize $\chi^0\!\! \neq \!
    \textsf{lsp}$} 
  \includegraphics[width=0.45\textwidth]{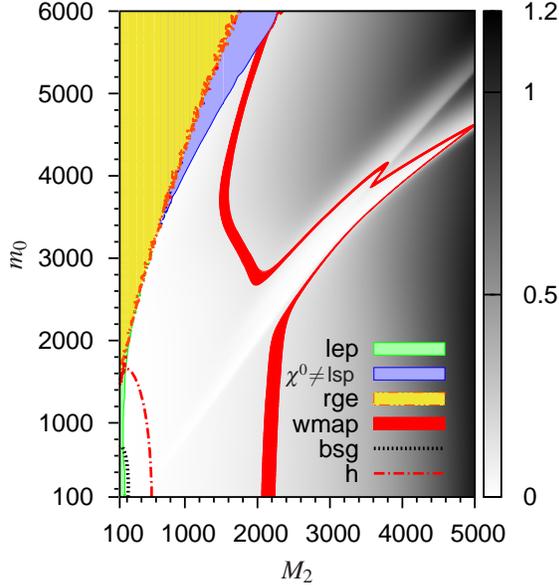}
  \caption{Relic density $\Omega_\chi h^2$ in the representation
    \textbf{210} and $H=$ \SU51 in ($M_{2}, m_0$) plane: $\tan\beta =
    10,\ \mathrm{sgn}(\mu) = +1,\ A_0=0$. The key is explained in
    the text.}
  \label{fig:relic210}
\end{figure}

In the representation \textbf{210} we inspected the breaking chain
through the intermediate gauge group \SU51 (called flipped SU(5)).
In Figure \ref{fig:relic210} the area of preferred thermal relic
density in the representation \textbf{210} is plotted for a set of
(GUT scale) parameters. For the chosen parameters, rather large \WMAP\
preferred regions are found for the large values of \M2 and/or \m0
parameters.
The dark shaded areas represent larger relic density than the lighter
areas.  \textsf{wmap} denoted filling is the \WMAP\ preferred region,
\textsf{lep} shows an area, where the experimental mass limits are not
met, \textsf{rge} shows an area where there is no radiative \EWSB, and
\textsf{lsp} an area where neutralino is not the LSP. For the relic
density, we use here the \WMAP\ combined three year limits
\cite{Spergel:2006hy}
\begin{eqnarray}
  \Omega_{CDM} h^2 = 0.11054^{+0.00976}_{-0.00956} \quad (2\sigma).
\end{eqnarray}
The tiny area enclosed by the \textsf{bsg} contour is disallowed by $b
\to s \gamma$ constraint.
We have used the two sigma world average of $BR(b\to s \gamma) = (355
\pm 24^{+9}_{-10} \pm 3) \times 10^{-6}$ for the branching fraction
\cite{Barberio:2007cr}. The dash-dotted line (\textsf{h}) encloses an
area with $m_h<114$ GeV.
The \textsf{lep} limits for the particle masses are the same as in
\cite{Huitu:2007vw}.  The spectrum was calculated with {\tt SOFTSUSY}
\cite{Allanach:2001kg} and relic densities and constraints with {\tt
  micrOMEGAs} \cite{Belanger:2006is}.

The \WMAP preferred region is very wide as compared to e.g. models in
typical mSUGRA or SU(5) \cite{Huitu:2008sa,Huitu:2007vw}. For a large
part of the parameter space the relic density is below the \WMAP\
limit.  The lightest neutralino is either wino or higgsino, which
explains the low overall relic density. In the diagonal of the figure
the lightest chargino mass is very close to the lightest neutralino
mass leading to the remarkable co-annihilation through the processes
$\Neu 1 \Cha 1 \to q_u\, \wbar q_d$ and $\Neu 1\Neu 1, \Cha 1 \Cha 1
\to q\, \wbar q, \ell \wbar \ell, W^+ W^-$. The spectrum is relatively
heavy for the \WMAP\ preferred region, but the value of $\mu$ stays
around a TeV, which doesn't necessarily require large fine-tuning.

\subsection{Representation 54}
The two chains of the {\bf 54} dimensional representation do not lead
to large areas with acceptable relic density.  Because $\Neu 1$ is
mostly bino in \SUSO7 chain, the spectrum with preferred RD is quite
light and conflicts with collider constraints in some parts of the
parameter space (although the higgsino component keeps the overall
thermal relic density lower than for example in the universal SUGRA
model).  In \SUPS the higgsino component in the lightest neutralino is
larger, and also co-annihilation with chargino is present at some
points of parameter space, so the allowed region extends also to
heavier spectrum. For more detailed analysis the reader is referred to
an upcoming publication \cite{teossa-so10}.

\thispagestyle{plain} 

\section{Summary}
\label{sec:summary}

We studied the dark matter allowed regions in the SO(10) GUT
representations, of which all but the singlet may lead to
non-universal gaugino masses.  The \WMAP\ preferred relic density
regions were quite distinct for different representations, thus
leading to quite different particle spectra for each representation.
As an example, we showed the behavior of the thermal relic density in
the representation {\bf 210}, since there the spectrum is quite heavy
as compared to the universal SUGRA case. It was also shown, that the
\WMAP\ preferred relic density area is very large in this case.
Moreover, it is important to realize that there is no automatical
theoretical preference for the gaugino masses to be unified.


\begin{theacknowledgments}
  The work of K.H.~is supported by the Academy of Finland (project
  No.~115032). The work of J.L.~is supported by the Bundesministerium
  f\"ur Bildung und Forschung, Berlin-Bonn.
\end{theacknowledgments}

\bibliographystyle{aipproc}
\bibliography{bib-relicdensity-so10.bib}

\end{document}